%% file: NeuroComp08.tex
\documentclass[onecolumn,a4,10pt,nofootinbib]{article}
\usepackage{iasted}
\usepackage{times}
\usepackage[dvips]{graphicx}
\usepackage{multicol}
\usepackage{bm}
\usepackage{amsfonts}
\usepackage{amssymb}
\usepackage{amsmath}

\input{macro}

\begin{document}

\input{title}
\begin{multicols}{2}
\input{introduction}

\input{framework}

\input{dyneffects}

\input{conclusion}
\bibliographystyle{unsrt0}
{\tiny  \bibliography{biblio}}

\end{multicols}
\begin{figure}[!ht]
\includegraphics[width=16cm,height=10cm]{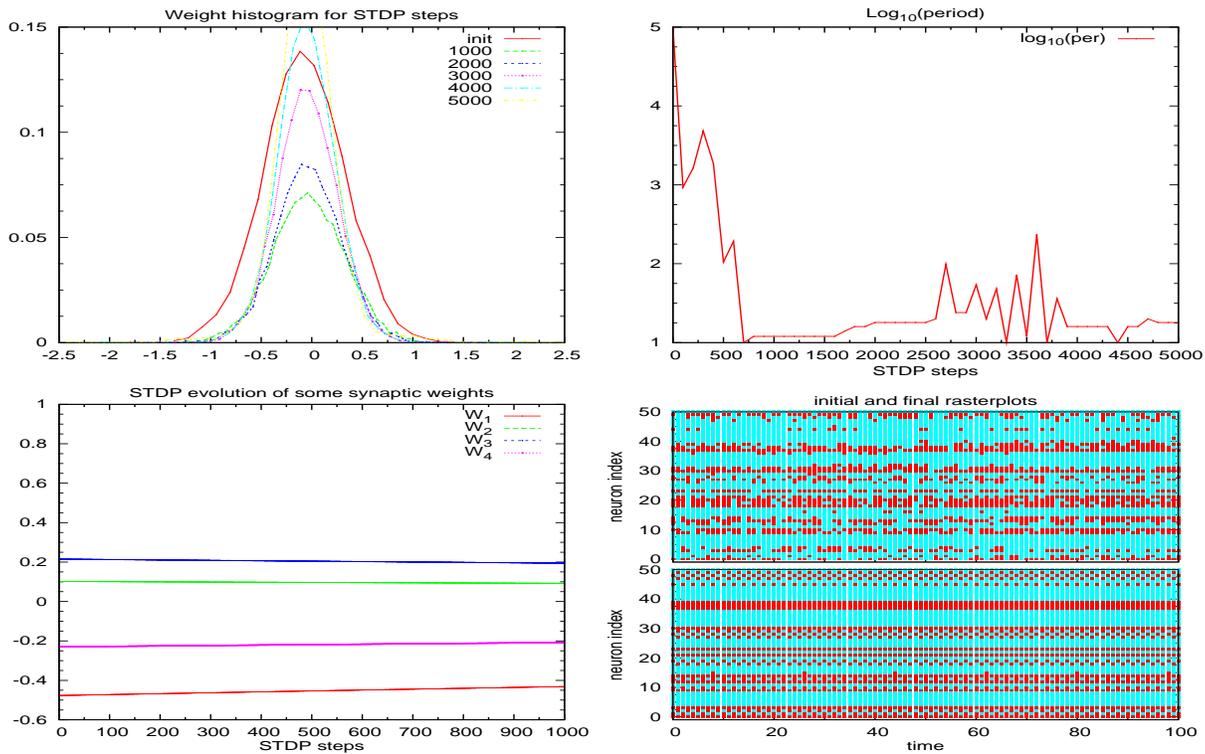}
\caption{Effect of synaptic plasticity on: (up left) synaptic weights histogram;
(up right) Period of the orbit; (down left) some synaptic weights evolution;
(down right) raster plot of a few neurons.   \label{Fex}}
\end{figure}

\end{document}

%% file: macro.tex
\newcommand{\bd}{\begin{document}}
\newcommand{\ed}{\end{document}}
\newcommand{\beq}{\begin{equation}}
\newcommand{\eeq}{\end{equation}}
\newcommand{\nid}{\noindent}
\newcommand{\ben}{\begin{enumerate}}
\newcommand{\een}{\end{enumerate}}
\newcommand{\bit}{\begin{itemize}}
\newcommand{\eit}{\end{itemize}}
\newcommand{\baR}{\begin{array}}
\newcommand{\eaR}{\end{array}}
\newcommand{\su}[1]{\section{#1}}
\newcommand{\ssu}[1]{\subsection{#1}}
\newcommand{\sssu}[1]{\subsubsection{#1}}
\newcommand{\deq}{\stackrel {\rm def}{=}}
\newcommand{\bea}{\begin{eqnarray}}
\newcommand{\eea}{\end{eqnarray}}
\newcommand{\SOS}[1]{\bf[#1]}
\newcommand{\st}[1]{{\nid\bf #1}}
\newcommand{\emp}[1]{{\it #1}}

\newcommand\D{\displaystyle}

\newcounter{CDef}[section]
\newtheorem{definition}{Definition}
\newcommand{\bdf}{\begin{definition}}
\newcommand{\edf}{\end{definition}}
\newcommand{\bpr} {\noindent {\bf Proof }}
\newcommand{\epr}{\rule{0.25cm}{0.25cm}\\}

\newtheorem{theorem}{Theorem}
\newcommand{\bth}{\begin{theorem}}
\newcommand{\enth}{\end{theorem}}
\newcounter{Cprop}[section]
\newtheorem{proposition}{Proposition}
\newcommand{\bp}{\begin{proposition}}
\newcommand{\ep}{\end{proposition}}
\newcounter{Ccor}\newtheorem{corrollary}{Corrollary}
\newcommand{\bcor}{\begin{corrollary}}
\newcommand{\ecor}{\end{corrollary}}
\newcounter{CLem}\newtheorem{lemma}{Lemma}
\newcommand{\blem}{\begin{lemma}}
\newcommand{\elem}{\end{lemma}}
\newcounter{Cconj}\newtheorem{conjecture}{Conjecture}
\newcommand{\bconj}{\begin{conjecture}}
\newcommand{\econj}{\end{conjecture}}
\newcommand\rep{\rightarrow}
\newcommand\drep{\downharpoonright}

\newcommand\bbbr{{\sf I\!R}}
\newcommand\bbbrn{{\sf I\!R}^N}
\newcommand\bbbn{{\sf I\!N}}

\newcommand\Vm{V_{min}}
\newcommand\VM{V_{max}}

\newcommand\ei{{\bf e}_i}
\newcommand\X{{\bf X}}
\newcommand\VTt{{\V^{(\tau)}(t)}}
\newcommand\VTtp{{\V^{(\tau)}(t+1)}}
\newcommand\Y{{\bf Y}}
\newcommand\F{{\bf F}}
\newcommand\Fg{{\F_{\bg}}}
\newcommand\FgT{{\F_{\gT}}}
\newcommand\Fgt{{\F_{\bsg}^t}}
\newcommand\Fgta{{\F_{\bsg}^{t\ast}}}
\newcommand\Fgn{{\F_{\bsg}^n}}
\newcommand\Wij{W_{ij}}
\newcommand\WijT{W_{ij}^{(\tau)}}
\newcommand\WijTp{W_{ij}^{(\tau+1)}}
\newcommand\dWijT{\delta \WijT}
\newcommand\poT{\pi_{\tom^{(\tau)}}}
\newcommand\poTp{\pi_{\tom^{(\tau+1)}}}

\newcommand\tX{\tilde{\X}}

\newcommand\Iei{I^{ext}_i}

\newcommand\cA{{\cal A}}
\newcommand\cB{{\cal B}}
\newcommand\cF{{\cal F}}
\newcommand\cFg{{\cF_{\bg}(\phi)}}
\newcommand\cFgp{{\cF_{\bg'}(\phi)}}
\newcommand\cP{{\cal P}}
\newcommand\cR{{\cal R}}

\newcommand\Po{{\cP_{\bom}}}
\newcommand\Pot{{\cP_{\bom(t)}}}
\newcommand\pTo{{\pi_{\tom}^{(T)}}}
\newcommand\pToT{{\pi_{\tom^{(\tau)}}^{(T)}}}
\newcommand\po{{\pi_{\tom}}}
\newcommand\pop{{\pi_{\tom'}}}
\newcommand\piij{{\pi_{\tom;ij}}}
\newcommand\mua{{\mu_{\cA}}}
\newcommand\mula{{\mu^{Leb}_{\cA}}}
\newcommand\muo{{\mu_{\tom}}}
\newcommand\mg{{\mu_{\bg}}}
\newcommand\mpg{{\nu_{\psi;\bg}}}

\newcommand\dT{{d_\Theta}}

\newcommand\BA{\cB(\cA)}
\newcommand\ba{\mbox{{\boldmath $\alpha$}}}
\newcommand\Bg{\mbox{{\boldmath $\Gamma$}}}
\newcommand\bg{\mbox{{\boldmath\tiny $\Gamma$}}}
\newcommand\gT{\mbox{{\boldmath $\Gamma$}}^{(\tau)}}
\newcommand\gTp{\mbox{{\boldmath $\Gamma$}}^{(\tau+1)}}
\newcommand\bsg{\mbox{{\boldmath\tiny $\gamma$}}}
\newcommand\bsl{\mbox{{\boldmath\tiny $\lambda$}}}
\newcommand\bl{\mbox{{\boldmath$\lambda$}}}
\newcommand\bls{\mbox{{\boldmath\tiny$\lambda^\ast$}}}
\newcommand\ble{\mbox{{\boldmath$\lambda^\ast$}}}
\newcommand\bphi{\mbox{{\boldmath$\phi$}}}

\newcommand\bom{\mbox{{\boldmath $\omega$}}}
\newcommand\tom{\tilde{\omega}}

\newcommand\sg{{\sigma_{\bg}}}
\newcommand\stg{{\sigma^t_{\bg}}}
\newcommand\stgo{{\sigma^t_{\bg}\tom}}

\newcommand\Slp{\left\{0,1\right\}^\bbbn}
\newcommand\Spg{\Sigma_{\bg}}
\newcommand\Spgp{\Sigma_{\bg'}}
\newcommand\Spgm{\Sigma_{\bg^{(m)}}}
\newcommand\SpgT{\Sigma_{\bg}^{(T)}}
\newcommand\Spgt{\Sigma_{\bg^{(\tau)}}}
\newcommand\Spgtp{\Sigma_{\bg^{(\tau+1)}}}
\newcommand\Minv{\cM^{inv}_{\bg}}
\newcommand\STpo{S^{(T)}_{\bg}\phi(\tom)}
\newcommand\STq{S^{(T)}_{\bg} \phi^{(4)}(.;\bl)}
\newcommand\pres{P_{\bg}(\psi)}
\newcommand\grad{\nabla_{\bg}}
\newcommand\gradPT{\nabla_{\bg}\PT}

\newcommand\psiT{\psi^{(\tau)}}
\newcommand\psiTp{\psi^{(\tau+1)}}
\newcommand\PT{P(\psiT)}
\newcommand\PTp{P(\psiTp)}
\newcommand\npT{\nu_{\psiT}}

\newcommand\muu{\nu^{(1)}_{\bl;\bg}}
\newcommand\Pu{P^{(1)}_{\bg}(\bl)}
\newcommand\pu{\phi^{(1)}(\tom;\bl)}
\newcommand\mud{\nu^{(2)}_{\bl;\bg}}
\newcommand\Pd{P^{(2)}_{\bg(\bl)}}
\newcommand\pd{\phi^{(2)}(\tom;\bl)}
\newcommand\mut{\nu^{(3)}_{\bl;\bg}}
\newcommand\Pt{P^{(3)}_{\bg}(\bl)}
\newcommand\pt{\phi^{(3)}(\tom;\bl)}
\newcommand\muq{\nu^{(4)}_{\bl;\bg}}
\newcommand\Pq{P^{(4)}_{\bg}(\bl)}
\newcommand\pq{\phi^{(4)}(\tom;\bl)}
\newcommand\psigo{\psi(\tom;\small{\bg})}
\newcommand\phigo{\phi(\tom;\small{\bg})}

\newcommand\nua{\nu_{\cA}}

\newcommand\gijo{g_{ij}^{(0)}(W_{ij})}
\newcommand\gijou{g_{ij}^{(01)}(W_{ij},\omejt)}
\newcommand\gijuo{g_{ij}^{(10)}(W_{ij},\omeit)}
\newcommand\gijdu{g_{ij}^{(21)}(W_{ij},\omeit^2\times \omejt)}
\newcommand\gijud{g_{ij}^{(12)}(W_{ij},\omeit\times \omejt^2)}
\newcommand\gijuu{g_{ij}^{(11)}(W_{ij},\omeit\times \omejt)}
\newcommand\gijkl{g_{ij}^{(kl)}(W_{ij},\omeitk\times \omejtl)}
\newcommand\hijkl{h_{ij}^{(kl)}(W_{ij})}

\newcommand\aijo{a_{ij}^{(0)}}
\newcommand\aijou{a_{ij}^{(01)}(\omejt)}
\newcommand\aijuo{a_{ij}^{(10)}(\omeit)}
\newcommand\aijdu{a_{ij}^{(21)}(\omeit^2\times \omejt)}
\newcommand\aijud{a_{ij}^{(12)}(\omeit\times \omejt^2)}
\newcommand\aijuu{a_{ij}^{(11)}(\omeit\times \omejt)}
\newcommand\aijkl{a_{ij}^{(kl)}(\omeitk\times \omejtl)}

\newcommand\aijtou{a_{ij}^{(01)}(W_{ij},\pTo(\tomej))}
\newcommand\aijtuo{a_{ij}^{(10)}(W_{ij},\pTo(\tomei))}
\newcommand\aijtdu{a_{ij}^{(21)}(W_{ij},\pTo(\tomei^2\times \tomej))}
\newcommand\aijtud{a_{ij}^{(12)}(W_{ij},\pTo(\tomei\times \tomej^2))}
\newcommand\aijtuu{a_{ij}^{(11)}(W_{ij},\pTo(\tomei\times \tomej))}

\newcommand\paijou{a_{ij}^{(01)}(W_{ij},\po(\omej))}
\newcommand\paijuo{a_{ij}^{(10)}(W_{ij},\po(\omei))}
\newcommand\paijdu{a_{ij}^{(21)}(W_{ij},\po(\tomei^2\times \tomej))}
\newcommand\paijud{a_{ij}^{(12)}(W_{ij},\po(\tomei\times \tomej^2))}
\newcommand\paijuu{a_{ij}^{(11)}(W_{ij},\po(\tomei\times \tomej))}
\newcommand\paijkl{a_{ij}^{(kl)}(W_{ij},\po(\tomeik\times \tomejl))}

\newcommand\CoT{[\bom(1)\bom(2) \dots \bom(T)]}
\newcommand\Catn{[\ba(t_1)\ba(t_2) \dots \ba(t_n)]}
\newcommand\omei{\omega_i}
\newcommand\omej{\omega_j}
\newcommand\tomei{\tom_i}
\newcommand\tomej{\tom_j}
\newcommand\tomeik{\tom_i^{(k)}}
\newcommand\tomejl{\tom_j^{(l)}}
\newcommand\omeit{\left[\omega_i\right]_{0,T}}
\newcommand\omejt{\left[\omega_j\right]_{0,T}}
\newcommand\omeitk{\left[\omega_i\right]^{k}_{t-T,t}}
\newcommand\omejtl{\left[\omega_j\right]^{l}_{t-T,t}}

\newcommand\ZTp{Z_{\bg}^{(T)}(\psi)}

\newcommand\Vr{V_{reset}}
\newcommand\V{{\bf V}}
\newcommand\hV{\hat{V}}

\newcommand\I{{\bf I}}
\newcommand\J{{\bf J}}
\newcommand\Ia{I^{(ion)}}
\newcommand\Iw{I^{(w)}}
\newcommand\ie{i^{(ext)}}
\newcommand\Ie{{\bf{I}}^{(ext)}}
\newcommand\Is{I^{(syn)}}
\newcommand\Isk{I^{(syn)}_k}
\newcommand\hIa{\hat{I}^{(ion)}}

\newcommand\tk{\tau^{(k)}}
\newcommand\tik{t_i^{(k)}}
\newcommand\til{t_i^{(l)}}
\newcommand\tjn{t_j^{(n)}}
\newcommand\tpm{\tau^{\pm}}
\newcommand\bG{\bar{G}}
\newcommand\cC{{\cal C}}
\newcommand\cD{{\cal D}}
\newcommand\cE{{\cal E}}
\newcommand\cG{{\cal G}}
\newcommand\cH{{\cal H}}
\newcommand\cI{{\cal I}}
\newcommand\cN{{\cal N}}
\newcommand\cM{{\cal M}}
\newcommand\cS{{\cal S}}
\newcommand\cU{{\cal U}}
\newcommand\cW{{\cal W}}
\newcommand\dW{\delta \cW}
\newcommand\cWT{\cW^{(\tau)}}
\newcommand\dWT{\delta\cW^{(\tau)}}
\newcommand\cWTp{\cW^{(\tau+1)}}
\newcommand\cWs{\cW^{(\ast)}}

\newcommand\Bev{\cB(\V,\epsilon)}
\newcommand\bmd{\cB_\cM(\delta)}

\newcommand\ton{\left[\tilde{\omega}\right]_n}
\newcommand\tot{\left[\tilde{\omega}\right]_t}
\newcommand\totm{\left[\tilde{\omega}\right]_{t-1}}
\newcommand\tos{\left[\tilde{\omega}\right]_s}
\newcommand\toT{\left[\tilde{\omega}\right]_T}
\newcommand\bomt{\bom(t)}
\newcommand\bdom{\bar{\cD}(\bom)}
\newcommand\com{\cC(\bom)}
\newcommand\comt{\cC(\bom(t))}
\newcommand\dom{\cD(\bom)}
\newcommand\mjt{M_j(t,\tom)}

\newcommand\cMo{{\cM_{\bom}}}
\newcommand\cMot{\cM_{\bom(t)}}
\newcommand\Mot{\cM_{\tot}}
\newcommand\Motm{\cM_{\totm}}
\newcommand\MoT{\cM_{\toT}}
\newcommand\Mos{\cM_{\bom(s)}}
\newcommand\oM{\Omega}

\newcommand\be{{\bf e}}
\newcommand\FT{\F^T}
\newcommand\FTp{\F^{T+1}}
\newcommand\bcF{\bar{\cF}}
\newcommand\Ft{\F\left[.;.,\tom\right]}
\newcommand\Fto{\F\left[.;t,\tom\right]}
\newcommand\Ftot{\F^{t+1}_{\tom}}
\newcommand\Ftso{\F^{t,s}_{\tom}}
\newcommand\Fkto{F_k\left(t,\tom\right)}
\newcommand\Fkso{F_k\left(s,\tom\right)}
\newcommand\FkVto{F_k\left(t,\tom\right)\V}
\newcommand\FVTo{\F^{t+1}_{\tom}(\V)}
\newcommand\FTo{\F^{t}_{\tom}}
\newcommand\FkTo{F^{t}_{k;\tom}}
\newcommand\DFFTV{\left. D\FTF\right|_{\V}}
\newcommand\DFFTpV{\left. D\FTpF\right|_{\V}}

\newcommand\dAS{d(\oM,\cS)}
\newcommand\dASe{d^{(exp)}(\oM,\cS)}

\newcommand\GFTe{\Gamma_{\cF,T,\epsilon}}
\newcommand\GFTpe{\Gamma_{\cF,{T+1},\epsilon}}
\newcommand\GFTep{\Gamma_{\cF,T,\epsilon'}}

\newcommand\PF{\Pi_{\cF}}
\newcommand\PbF{\Pi_{\bcF}}
\newcommand\VF{{\V_{\cF}}}
\newcommand\VFb{{\V_{\bcF}}}
\newcommand\FF{{\F_{\cF}}}
\newcommand\FTF{{\FT_{\cF}}}
\newcommand\FTpF{{\FTp_{\cF}}}
\newcommand\FFb{{\F_{\bcF}}}
\newcommand\oGF{{\Omega(\GF)}}
\newcommand\oGFTe{{\Omega(\GFTe)}}
\newcommand\oFTe{{\Omega_{\cF}(\GFTe)}}
\newcommand\cPT{\cP^{(T)}}

\newcommand\Jkto{J_k(t,\tot)}

\newcommand\gkt{\gamma_k(t,\tot)}
\newcommand\gks{\gamma_k(s,\tos)}
\newcommand\skt{\sigma_k(t,\tot)}

\newcommand\heff{h^{(eff)}}
\newcommand\PrW{\cP_\cW}
\newcommand\PrI{\cP_{\Ie}}
\newcommand\PrWi{\cP_{\cW,\Ie}}

\newcommand\ltjn{\left\{\tjn\right\}_t}

\newcommand\wsg{\cW^s(\V)}
\newcommand\wsl{\cW^s_{loc}(\V)}
\newcommand\wslf{\cW^s_{loc}(\F(\V))}
\newcommand\wse{\cW^s_{\epsilon}(\V)}
\newcommand\wset{\cW^s_{\epsilon}(\V(t))}
\newcommand\psis{\psi^\ast}
\newcommand\nus{\nu^\ast}
\newcommand\bgs{\bg^\ast}

%% file: title.tex
\title{Statistics of spikes trains, synaptic plasticity and Gibbs distributions.}

\author{
B. Cessac
\thanks{INRIA, 2004 Route des Lucioles, 06902 Sophia-Antipolis, France.}
\thanks{Laboratoire Jean-Alexandre Dieudonn\'e, U.M.R. C.N.R.S. N° 6621, Nice, France.}
\thanks{Universit\'e de Nice, Parc Valrose, 06000 Nice, France.},
H. Rostro $^\ast$,
J.C. Vasquez $^\ast$,
T. Vi\'eville $^\ast$
}

\date{\today}

\maketitle

\begin{abstract}
We introduce a mathematical
framework where the statistics of spikes trains, produced by neural networks
 evolving under synaptic plasticity, can be analysed. 
\end{abstract}

%% file: introduction.tex
\su{Introduction.} 

 Synaptic plasticity occurs at many levels of organization and time scales in the nervous system 
\cite{bienenstock-et-al:82}. 
It is of course
involved in memory and learning mechanisms, but it also alters excitability of brain area
and regulates behavioural states (e.g. transition between sleep and wakeful activity).
Therefore, understanding the effects of synaptic plasticity on neurons dynamics
is a crucial challenge. However, the exact relation between the synaptic properties (``microscopic'' level) 
and the  effects induced on neurons dynamics (meso-or macroscopic level) is still highly controversial.

On experimental grounds, synaptic changes can be induced by \textit{specific} simulations conditions
defined through  the firing frequency of pre- and postsynaptic
neurons \cite{bliss-gardner:73,dudek-bear:93}, the membrane potential of the postsynaptic
neuron \cite{artola-et-al:90}, spike timing \cite{levy-stewart:83,markram-et-al:97,Bi-Poo:01}
(see \cite{malenka-nicoll:99} for a review). Different mechanisms have been exhibited from the Hebbian's ones \cite{hebb:49} to Long Term
Potentiation (LTP) and Long Term Depression (LTD), and more recently to Spike Time Dependent
Plasticity (STDP) \cite{markram-et-al:97,Bi-Poo:01} (see \cite{dayan-abbott:01,gerstner-kistler:02,cooper-et-al:04}
for a review). Most often, these simulation are performed in isolated
neurons in \textit{in vitro} conditions. Extrapolating the action of these mechanisms to
in vivo neural networks requires both a bottom-up and up-bottom approach.

 This issue is
tackled, on theoretical grounds, by infering ``synaptic updates rules'' or ``learning rules'' from
biological observations \cite{vondermalsburg:73,bienenstock-et-al:82,miller-et-al:89}
 (see \cite{dayan-abbott:01,gerstner-kistler:02,cooper-et-al:04}
for a review) and extrapolating, by theoretical or numerical investigations, what are the effects
of such synaptic rule on such neural network \textit{model}. This approach relies on the belief that
there are ``canonical neural models'' and ``canonical plasticity rules'' capturing the most essential
features of biology. Unfortunately, this results in a plethora of canonical ``candidates'' and
a huge number of papers and controversies. Especially, STDP deserved a long discussion either on its biological
interpretation or on its practical implementation \cite{izhikevich-desai:03}. 
Also, the discussion of  what is relevant for neuron coding and what has to be measured to conclude
on the effects of synaptic plasticity is a matter of long debates (rate coding, rank coding, spike coding ?) \cite{cessac-et-al:08,rieke-warland:96}

 In an attempt to clarify and unify the overall vision, some researchers have proposed to associate learning rules
and their dynamical effects to general principles, and especially to``variational'' or ``optimality'' principles, where some functional
has to maximised or minimised (known examples of variational principles in 
physics are least-action  or entropy maximization).
Dayan and Hausser \cite{dayan-hausser:04} have shown that STPD can be  viewed as an optimal noise-removal filter for certain
noise distributions, Rao and Sejnowski \cite{rao-sejnowski:99,rao-sejnowski:01} suggested that STDP may be related to
optimal prediction (a neuron attemps to predict its membrane potential at some time, given
the past). Bohte and Mozer \cite{bohte-mozer:07}
proposes that STPD minimizes response variability.  Chechik \cite{chechik:03} relates STDP to information theory via maximisation of mutual
information between input and output spike trains. In the same spirit,  Toyoizumi et al \cite{toyoizumi-et-al:05,toyoizumi-et-al:07}
 have proposed to associate STDP to an optimality principle where transmission of information
between an ensemble of presynaptic spike trains and the ouput spike train of the postsynaptic neuron is optimal,
 given some constraints imposed by biology (such as homeostasy and minimisation in the number of strong synapses, which
is costly in view of continued protein synthesis). ``Transmission of information'' is measured 
by the mutual information for  the joint probability distribution of the input and output spike trains. 
Therefore, in these ``up-bottom'' approaches, plasticity rules ``emerge'' from first principles. 

Obviously, the validations of these theories requires a model of neurons dynamics and
 a statistical model for  spike trains. Unfortunately, often only isolated neurons are considered, submitted to input 
spike
trains with ad hoc statistics (typically, Poisson distributed with independent spikes  \cite{toyoizumi-et-al:05,toyoizumi-et-al:07}).
However, adressing the effect of synaptic plasticity in neural networks where dynamics is \textit{emerging} 
from collective effects and where spike statistics are \textit{constrained} by this dynamics
seems  to be of central importance. 

This issue is subject to two main difficulties.
On the one hand, one must identity the generic dynamical regimes displayed by a neural network
model for different choices of parameters (including synaptic strength). On the other hand,
one must analyse the effects of varying synaptic weights when applying plasticity rules.
This requires to handle a complex interwoven evolution where neurons dynamics depends on synapses
and synapses evolution depends on neuron dynamics. The first aspect has been adressed by several 
authors using mean-field approaches (see \cite{samuelides-cessac:07} and references therein), ``markovian approaches''
\cite{chow-soula:07}, or dynamical system theory (see \cite{cessac-samuelides:07} and references therein).
The second aspect has, to the best of our knowledge, been investigated theoretically in only a few
examples with hebbian learning \cite{dauce-et-al:98,siri-et-al:07,siri-et-al:08} or discrete
time Integrate and Fire models with an STDP like rule \cite{soula:05,soula-et-al:06}. 

Following these tracks, we have investigated the dynamical effects of a subclass of synaptic plasticity
rules (including some implementations of STDP) in neural networks models where one has a full characterization
of the generic dynamics  \cite{cessac:08,cessac-vieville:08}. Thus, our aim is not to provide general statements
about synaptic plasticity in biological neural networks. We simply want to have a good mathematical control
of what is going on in specific models, with the hope that this analysis should some light on what happens
(or  \textit{does not} happen) in ``real world'' neural systems.  
Using the framework of ergodic theory and thermodynamic
formalism, these plasticity rules can be formulated as a variational principle for a quantity, called the topological
pressure, closely related to thermodynamic potentials, like free energy or Gibbs potential in statistical physics \cite{keller:98}. 
  As a main consequence of this formalism the statistics of spikes
are more likely described by a Gibbs probability distributions
than by the classical Poisson distribution. 

In this communication,  we give a brief outline of this setting and provide a simple illustration.
Further developments will be published in an extended paper.

%% file: framework.tex
\su{General framework.}
\st{Neuron dynamics.} In this paper we consider a simple implementation of the leaky
 Integrate and Fire model, where time has been discretized. This model
has been introduced and analysed by mean-field technics in \cite{soula-et-al:06}.
A full characterisation of its dynamics has been done in \cite{cessac:08}
and most of the conclusions extend to discrete time Integrate and Fire models with adaptive conductances
\cite{cessac-vieville:08}.

 Call $V_i$ the membrane potential of neuron $i$. 
Fix a firing threshold
$\theta>0$. Then the dynamics is given by: 
\beq \label{DNN}
V_i(t+1)=
\gamma V_i(t) \left(1 - Z[V_i(t)] \right)+ \sum_{j=1}^N W_{ij}Z[V_j(t)]+ \Iei,
\eeq
\nid  $i=1 \dots N$, where  the ``leak rate'' $\gamma \in [0,1[$ and $Z(x)=\chi(x \geq \theta)$ where $\chi$ is the indicatrix function namely,
$Z(x)=1$ whenever $x \geq \theta$ and $Z(x)=0$ otherwise. $W_{ij}$ models the synaptic weight from $j$ to $i$.
Denote by $\cW$ the matrix of synaptic weights. $\Iei$ is some (time independent) external current. Call $\Ie$
the vector of $\Iei$. $\cW,\Ie$ are control parameters for the dynamics.
To alleviate the notation we write 
$\Bg = \left( \cW,\Ie \right)$.
Thus $\Bg$ is a point in a $N^2+N$ dimensional space of control parameters.

It has been proved in \cite{cessac:08} that the dynamics of (\ref{DNN}) admits
generically finitely many periodic orbits
\footnote{
As a side remark, we would 
like to point out that this models admits generically a finite Markov partition,
which allows to describe the evolution of probability distributions via a Markov chain. This gives some support to
the approach developped in \cite{chow-soula:07} though, our analysis also shows, in the present example,
 that, in opposition to what is assumed in \cite{chow-soula:07}, 
(i) the size of the Markov partition (and especially its finitness) does not only depend on membrane and synaptic time constants,
but on the values of the synaptic weights; 
(ii) the probability that, exactly $j$ neurons in the network
fire and exactly $N-j$ neurons does not fire at a given time, does not factorize 
(see eq. (1) in \cite{chow-soula:07});
(iii) the invariant measure is not unique.
}. However, in some region of the parameters
space, the period of these orbits can be quite larger than any numerically accessible time,
leading to a regime practically indistinguishable from chaos. \\

\st{Spikes dynamics.} 
Provided that synaptic weights and currents are bounded, it is easy to show
that the trajectories of (\ref{DNN}) stay in a compact set 
 $\cM=[\Vm,\VM]^N$. For each neuron one can decompose the interval $\cI = [\Vm,\VM]$ into
$\cI_0 \cup \cI_1$ with $\cI_0=[\Vm,\ \theta[$, $\cI_1=[\theta,\VM]$. If
 $V \in \cI_0$ the neuron is quiescent, otherwise it fires.
This splitting induces a partition $\cP$ of $\cM$, that we call the ``natural partition''.
The elements of $\cP$ have the following form. 
 Call $\Lambda=\left\{0,1\right\}^N$,
$\bom=\left[\omega_i\right]_{i=1}^N \in \Lambda$. Then, $\cP=\left\{\Po \right\}_{\omega \in \Lambda}$,
where $\Po = \cP_{\omega_1} \times \cP_{\omega_2} \times \dots \times \cP_{\omega_N}$.
Equivalently, if $\V \in \Po$, then all neurons such that $\omega_i=1$ are firing
while neurons such that $\omega_i=0$ are quiescent. 
We call therefore $\bom$ a  ``spiking pattern''.
To each initial condition $\V \in \cM$
 we associate a ``raster plot'' $\tom=\left\{\bom(t)\right\}_{t=0}^{+\infty}$
such that $\V(t) \in \Pot, \forall t \geq 0$. We write $\V \rep \tom$.
Thus, $\tom$ is the sequence of spiking patterns
displayed by the neural network when prepared with the initial condition $\V$.
We denote by $\left[\omega\right]_{s,t}$ the sequence $\bom(s),\bom(s+1), \dots, \bom(t)$ of spiking patterns.
We say that an infinite sequence $\tom=\left\{\bom(t)\right\}_{t=0}^{+\infty}$
is \textit{an admissible raster plot} if there exists $\V \in \cM$ such that   
$\V \rep \tom$. We call $\Spg$ the set of admissible raster plots for the 
 parameters $\Bg$. The dynamics  on the set of raster plots, induced by (\ref{DNN}),
is simply the left shift $\sg$ (i.e. $\tom'=\sg \tom$ is such that $\bom'(t)=\bom(t+1), t \geq 0$).
Note that there is generically a one to one correspondence between the orbits of (\ref{DNN})
and the raster plots (i.e. raster plots provide a symbolic coding for the orbits) \cite{cessac:08}.  \\

\st{Statistical properties of orbits.}
We are interested in the statistical properties
of raster plots which are inherited from the statistical properties of orbits
of (\ref{DNN})  via the correspondence $\nu[A]=
\mu\left[\left\{\V \rep \tom, \tom \in A  \right\}\right]$,
where $\nu$ is a probability measure on $\Spg$, $A$ a measurable
set (typically a  cylinder set) and $\mu$ an \emp{ergodic} measure on $\cM$.
$\nu(A)$ can be estimated by the empirical average $\pTo(A)=  \frac{1}{T}\sum_{t=1}^T \chi(\stgo \in A)$
where $\chi$ is the indicatrix function. We are seeking
asymptotic statistical properties corresponding to taking
the limit $T \to \infty$. Let :

\beq\label{defpo}
\po(.)=\lim_{T \to \infty} \pTo(.),
\eeq

\nid be the \emp{empirical measure}  for the raster plot $\tom$, then $\nu=\po$ for $\nu$-almost every $\tom$. \\

\st{Gibbs measures.}
The explicit form of $\nu$ is not known in general.
Moreover, one has experimentally only access 
to finite time raster plots and
the convergence to $\po$ in (\ref{defpo})
can be quite slow. Thus, it is necessary
to provide some \emp{a priori} form for the probability
$\nu$. We call a \emp{statistical model}
 an ergodic probability measure $\nu'$ which can serve as a prototype\footnote{e.g. by minimising the Kullback-Leiber divergence
between the empirical measure and $\nu'$.}
for $\nu$. Since there are many  ergodic measure,
there is not a unique choice for the statistical model.
Note also that it depends on the parameters $\Bg$. We 
 want to define a procedure allowing
one to select a statistical model.

 A \emp{Gibbs measure with potential\footnote{Fix $0 < \Theta < 1$.  Define a metric on $\Spg$ by $\dT(\tom,\tom')=\Theta^p$,
where $p$ is the largest integer such that $\bom(t)=\bom(t'), 1 \leq t \leq p-1$.
For a continuous function $\psi : \Spg \to \bbbr$ and $n \geq 1$ define
$var_n\psi = \sup \left\{ |\psi(\tom) - \psi(\tom')| : \bom(t)=\bom'(t), 1 \leq i \leq n  \right\}$.
$\psi$ is called a \emp{potential} if $var_n(\psi) \leq C \Theta^n, n=1,2 \dots$, where $C$
is some positive constant \cite{keller:98}.} $\psi$} is an ergodic measure $\mpg$ such that
$\mpg\left(\left[\omega\right]_{1,T}\right) \sim \frac{\exp \sum_{t=1}^T \psi(\stg \tom)}{\ZTp}$
where  $\ZTp=\sum_{\SpgT} \exp\sum_{t=1}^T \psi(\stg \tom) $, $\SpgT$ being 
the set of admissible spiking patterns sequences  of length $T$. Note that for the forthcoming discussion
it is important to make explicit the dependence of these quantities in the parameters $\Bg$.

 The \emp{topological pressure}\footnote{This quantity is analogous to thermodynamic potentials  in statistical mechanics
like free energy or the grand potential $\Omega=U-TS-\mu N=-P V$, where $P$ is the thermodynamic pressure. Equation
(\ref{dP}) expresses that $\pres$ is the generating function for the cumùulants of the probability distribution.
Equation (\ref{supmu}) relates Gibbs measures to equilibrium states (entropy maximisation under constraints). }
 is the limit $\pres = \lim_{T \to \infty} \frac{1}{T} \log(\ZTp$.
 If $\phi$ is another potential one has :

\beq\label{dP}
\left.\frac{\partial  P_{\bg}(\psi +\alpha \phi)}{\partial \alpha}\right|_{\alpha=0}=\mpg[\phi],
\eeq

\nid the average value of $\phi$ with respect to $\mpg$.
The Gibbs measure obeys \textit{a variational principle}. 
Let $\nu$ be a $\sg$-invariant measure. Call $h(\nu)$ the entropy of  $\nu$. Let
$\Minv$ be the set of invariant measures for $\sg$, then:
\beq \label{supmu}
\pres=\sup_{\nu \in \Minv} \left( h(\nu)+\nu(\psi) \right)=h(\mpg)+\mpg(\psi).
\eeq

\medskip

\st{Adaptation rules.} We consider adaptation mechanisms where synaptic
weights evolve in time according to the spikes emitted 
by the pre- and post- synaptic neuron. The variation of $W_{ij}$ at time $t$ is a function
of the spiking sequences of neurons $i$ and $j$ from time $t-T$ to
time $t$, where $T$ is time scale characterizing the width of the spike
trains influencing the synaptic change. In this paper we investigate the effects of synaptic plasticity rules where the characteristic
time scale $T$ is quite a bit larger than the time scale of evolution of the neurons.

On practical/numerical  grounds we proceed as follows.
Let the neurons evolve over a time windows of width $T$, called an ``adaptation epoch''
during which synaptic weights are constant, and record the spikes trains.
From this record, we update the synaptic weights and a new 
adaptation epoch begins. We denote by $t$ the update index of neuron states 
 inside an adaptation epoch, while $\tau$ indicates the update index
of synaptic weights. Neuronal time is reset at each
new adaptation epoch.

As an example
we consider here an adaptation rule of type:
\beq  \label{STDPFC}
\baR{lll}
\dWijT= \WijTp-\WijT=\\
\epsilon
\left[
(\lambda-1) \WijT+\frac{1}{T-2C}g_{ij}(\omeit,\omejt)
\right].
\eaR
\eeq
The first term parametrized by $\lambda \in [0,1]$ mimics passive LTD
while:
$$g_{ij}(\omeit,\omejt)= \sum_{t=C}^{T-C} \sum_{u=-C}^{C} f(u)\omei(t+u) \omej(t),$$
\nid with :
\beq\label{fSTDP}
f(x)=
\left\{
\baR{lll}
A_- e^{\frac{x}{\tau_-}}, \ &x <0;\\
A_+ e^{-\frac{x}{\tau_+}}, \ &x >0;\\
0, \ &x=0,
\eaR
\right.
\eeq 
and with $A_-<0, A_+>0$, provides an example of STDP implementation.
Note that since $f(x)$ becomes negligible as soon as $x \gg \tau_+$ or $x \ll -\tau_-$, we
may consider that $f(u) =0$ whenever $|u| > C \deq 2 \max(\tau_+,\tau_-)$.
The parameter $\epsilon>0$ is chosen small enough to ensure adiabatic changes 
while updating synaptic weights. 

Updating the synaptic weights  has several consequences.
It obviously modifies the structure of the synaptic graph
with an impact on neurons dynamics.
But, it can also modify the statistics of spikes trains and
the structure of the set of admissible raster plots.
Let us now discuss these effects within more details. \\

%% file: dyneffects.tex
\st{Dynamical effects of synaptic plasticity.} Set $S_i(\tom)= \sum_{u=-C}^{C} f(u)\omei(u)$
and $H_{ij}(\tom)=\omega_j(0)S_i(\tom)$. Then the adaptation rule writes, as $T \to \infty$,
in terms of the empirical measure  $\poT$:
\beq\label{ruleth1}
\dWijT=\epsilon
\left(
(\lambda-1) \WijT+   \poT\left[H_{ij}\right]
\right),
\eeq
\nid where $\tom^{(\tau)}$ is the raster plot produced
by neurons in the adaptation epoch $\tau$. 

The term $S_i(\tom)$ can be either negative, inducing LTD or positive inducing LTP.
In particular, its average with respect to the empirical measure $\poT$ reads:
\beq \label{PoTSi}
\poT(S_i)=
\beta r_i(\tau) 
\eeq
\nid
where:
\beq
\beta=
\left[
A_- e^{-\frac{1}{\tau_-}}\frac{1-e^{-\frac{C}{\tau_-}}}{1-e^{-\frac{1}{\tau_-}}}
+
A_+ e^{-\frac{1}{\tau_+}}\frac{1-e^{-\frac{C}{\tau_+}}}{1-e^{-\frac{1}{\tau_+}}}.
\right]
\eeq
\nid and where $r_i(\tau)=\poT(\omei)$ is the frequency rate of neuron $i$ in the $\tau$-th adaptation epoch. 

The term $\beta$ neither depend on $\tom$ nor on $\tau$,
but only on the adaptation rule parameters $A_-,A_+,\tau_-,\tau_+,C$.
Equation (\ref{PoTSi}) leads 3 regimes.

\bit
\item\textbf{Cooperative regime.} If $ \beta>0$ then $\poT(S_i)>0$. Then synaptic weights have a tendency to become more positive.
 This corresponds to a cooperative system \cite{hirsch:89}. 
In this case,  dynamics become trivial with neurons firing at each time step or
remaining quiescent forever.

\item\textbf{Competitive regime.} On the opposite if $ \beta<0$  synaptic weights  become  negative.   
This corresponds to a competitive system \cite{hirsch:89}. 
In this case,  neurons remain quiescent forever.

\item\textbf{Intermediate regime.} The intermediate regime corresponds to $ \beta \sim 0$.
Here no clear cut tendency can be distinguished from the average value of $S_i$
and spikes correlations  have to be considered as well.
 In this situation, the effects of synaptic plasticity on
neurons dynamics strongly depend on the values of the parameter $\lambda$.
A detailed discussion of these effects will be published elsewhere.
 We simply
provide an example in Fig. \ref{Fex}. 
\eit

\st{Variational form.}
We now investigate the effects of the adaptation rule (\ref{ruleth1})
on the statistical distribution of the raster plot. To
this purpose we use, as a statistical model for the spike train distribution
in the adaptation epoch $\tau$, a Gibbs distribution
$\npT$ with a potential $\psiT$  and a topological pressure
$\PT$. Thus, the adaptation dynamics results in
a sequence  of  empirical measures $\left\{\poT \right\}_{\tau \geq 0}$, and
corresponding  statistical models $\left\{\npT \right\}_{\tau \geq 0}$,
corresponding to changes in the statistical properties of raster plots. 

These changes can be smooth, in which case  the
average value of observables changes smoothly.
Then,  using (\ref{dP}) 
we may write the adaptation rule (\ref{ruleth1}) in the form:
\beq\label{dWvar}
\baR{lll}
\dWT=\npT\left[\bphi \right]\\
=\nabla_{\tiny{\cW=\cWT}} P\left[\psiT+ (\cW-\cWT).\bphi(\cWT,.) \right]
\eaR,
\eeq
\nid where $\bphi$ is the matrix of components $\phi_{ij}$ with 
$\phi_{ij} \equiv \phi_{ij}(\Wij,\tom)=\epsilon\left((\lambda-1) \Wij+  H_{ij}\right)$
and where we use the notation
$(\cW-\cWT).\bphi=\sum_{i,j=1}^N (\Wij-\WijT) \phi_{ij}$.

This amounts to slightly perturb the potential $\psiT$ with a perturbation $(\cW-\cWT).\bphi$.
In this case, for sufficiently small $\epsilon$ in the adaptation rule, the variation of pressure between epoch $\tau$ and $\tau+1$ is given by:
$$\delta P^{(\tau)} \sim
\dWT.\npT\left[\bphi \right]=\dWT.\dWT >0
$$
Therefore the adaptation rule  is a \emp{gradient system which tends to \emp{maximize} the topological pressure}.

The effects of modifying synaptic weights can also
be sharp (corresponding typically to bifurcations in (\ref{DNN})).
We observe in fact 
phases of regular variations interrupted by singular transitions (see fig. \ref{Fex}).
These transitions have an interesting interpretation. 
When they happen the set of admissible raster plots is typically suddenly
modified by the adaptation rule. Thus, the set of admissible
raster plots obtained after adaptation
belongs to $\Spgt \cap \Spgtp$.
In this way, adaptation plays the role of a \emp{selective mechanism}
where the set of admissible raster plots, viewed as a neural
code, is gradually reducing, producing
after $n$ steps of adaptation a 
set $\cap_{\tau=1}^n \Spgt$ which can be 
 small (but not empty). 

If we consider the situation where (\ref{DNN}) is
a neural network submitted to some stimulus, where
a raster plot $\tom$ encodes the spike response to the stimulus
when the system was prepared in the initial condition
$\X \rep \tom$, then $\Spg$ is the set of all
possible raster plots encoding this stimulus.
This illustrates how daptation results in a reduction of the possible
coding, thus reducing the variability in the possible
responses. This property has also been observed in  \cite{dauce-et-al:98,soula-et-al:06,siri-et-al:07,siri-et-al:08}\\

\st{Equilibrium distribution.}
An ``equilibrium point'' is defined by $\dW=0$ and it corresponds to a maximum of the pressure
with :
$$\nus\left[\bphi \right]=0,$$
\nid where $\nus$ is a Gibbs measure with potential $\psis$.
This implies:
\beq
\cW^\ast=\frac{1}{1-\lambda} \nus\left[\cH \right]
\eeq
which gives, component by component, and making $\cH$ explicit:

\beq
W_{ij}^\ast=\frac{1}{1-\lambda} \sum_{u=-C}^{C} f(u) \nus\left[ \omej(0) \omei(u)\right],
\eeq

\nid where $\nus\left[ \omej(0) \omei(u)\right]$ is the probability, in the asymptotic regime,
that neuron $i$ fires at time $u$ and neuron $j$ fires at time $0$. Hence, in this case,
  the  synaptic
weights are purely expressed in terms of \emp{neurons pairs correlations}. As
a further consequence the equilibrium is described by a Gibbs distribution with
 potential $\bphi$. 

Therefore, if  the adaptation rule (\ref{STDPFC}) converges\footnote{This convergence
is ensured for sufficiently small $\epsilon$.} to an equilibrium point
as $\tau \to \infty$, it leads
naturally the system to a dynamics where raster plots are distributed
according to Gibbs distribution, with a pair potential having some analogy 
with an Ising model Hamiltonian. 
This makes  a very interesting link with the work
of Schneidman et al
\cite{schneidman-et-al:06} where
they show, using theoretical arguments as well as
empirical evidences on
experimental data for the salamander retina,
that spike trains statistics is likely described by a Gibbs distribution  with  an ``Ising''
like potential (see also \cite{tkacik-et-al:06}).

%% file: conclusion.tex
\su{Conclusion.} 

In this paper, we have introduced a mathematical
framework where spikes trains statistics, produced by neural networks,
possibly evolving under synaptic plasticity, can be analysed. 
It is argued that Gibbs distribution, arising naturally in ergodic theory,
are good candidates to provide efficient statistical models
for raster plots distribution. We have also shown that 
some plasticity rules can be associated to a variational
principle. Though only one example was proposed, it is easy
to extend this result to more general examples of adaptation rules.
This will be the subject of an extended forthcoming paper.

Concerning neuroscience there are several expected outcomes.
The approach proposed here is similar in spirit to the variational
approaches discussed in the introduction (especially 
\cite{chechik:03,toyoizumi-et-al:05,toyoizumi-et-al:07}).
Actually, the quantity ${\cal L}$ minimized in \cite{toyoizumi-et-al:05,toyoizumi-et-al:07}
can be obtained, in the present setting, with a suitable choice of potential.
But the present formalism concerns neural networks with intrinsic dynamics
(instead of isolated neurons submitted to uncorellated Poisson spikes trains).
Also, more general models could be taken into account (for example Integrate and Fire models 
with adaptive conductances \cite{cessac-vieville:08}). 

A second issue concerns spike trains statistics. As claimed by many authors and nicely
proved by Schneidman and collaborators \cite{schneidman-et-al:06,tkacik-et-al:06}, more elaborated
statistical models than uncorrelated Poisson distributions have to be considered to analyse spike trains statistics, especially taking
into account correlations between spikes \cite{nirenberg-latham:03}, but also higher order cumulants.
The present work show that Gibbs distributions, obtained from statistical inference in real data
by Schneidman and collaborators, may naturally arise from synaptic adaptation mechanisms. 
We also obtain an explicit form for the probability distribution depending e.g. on physical parameters
such as the time constants $\tau_+,\tau_-$ or the LTD/LTP strength $A_+,A_-$ appearing in
the STDP rule. This can be compared to real data. Also, having a ``good'' statistical
model is a first step to be able to ``read the neural code'' in the spirit of \cite{rieke-warland:96}, namely infer the conditional
probability that a stimulus has been applied given the observed spike train, knowing
the conditional
probability that one observes a spike train given the stimulus.

Finally, as a last outcome, this approach opens up the possibility of obtained
a specific spike train statistics from a deterministic neurons evolution
with a suitable synaptic plasticity rule (constrained e.g. by the potential $\bphi$).